\pdfoutput=1
\documentclass[conference, 9pt]{IEEEtran}
\IEEEoverridecommandlockouts
\usepackage{amsmath,amssymb,amsfonts}
\usepackage{algorithm}
\usepackage[noend]{algpseudocode}
\usepackage{graphicx}
\usepackage{float}
\usepackage{soul}
\usepackage{textcomp}
\usepackage{xcolor}
\usepackage{comment}
\usepackage{paralist}
\usepackage{wrapfig,lipsum,booktabs}
\usepackage{multirow}
\usepackage{color}
\usepackage{color, colortbl}
\usepackage{listings,multicol}
\usepackage{subcaption}
\newtheorem{theorem}{\textbf{Theorem}}
\newtheorem{definition}{\textbf{Definition}}

\newcommand{\mc}[1]{\mathcal{#1}}

\newcommand{\sd}[1]{{\color{red}{SD: {#1}}}}

\def\BibTeX{{\rm B\kern-.05em{\sc i\kern-.025em b}\kern-.08em
		T\kern-.1667em\lower.7ex\hbox{E}\kern-.125emX}}

\makeatletter
\def\endthebibliography{%
	\def\@noitemerr{\@latex@warning{Empty `thebibliography' environment}}%
	\endlist
}
\makeatother

\begin{document}
	\title{An RL-Based Adaptive Detection Strategy to Secure Cyber-Physical Systems}
%	\title{Catch Me If You Learn: A Reinforcement Learning Based Adaptive Monitoring Strategy for CPS
		% : An RL-Based Secure Control Strategy
		% \\{\footnotesize \textsuperscript{*}Note: Sub-titles are not captured in Xplore and
	
 	\author{\IEEEauthorblockN{1\textsuperscript{st} Ipsita Koley}
 		\IEEEauthorblockA{\textit{Dept. of Computer Science and Engineering} \\
 			\textit{Indian Institute of Technology}\\
 			Kharagpur, India\\
 			ipsitakoley@iitkgp.ac.in}
 		\and
 		\IEEEauthorblockN{2\textsuperscript{nd} Sunandan Adhikary}
 		\IEEEauthorblockA{\textit{Dept. of Computer Science and Engineering} \\
 			\textit{Indian Institute of Technology}\\
 			Kharagpur, India\\
 			sunandana@iitkgp.ac.in}
 		\and
 		\IEEEauthorblockN{3\textsuperscript{rd} Soumyajit Dey}
 		\IEEEauthorblockA{\textit{Dept. of Computer Science and Engineering} \\
 			\textit{Indian Institute of Technology}\\
 			Kharagpur, India\\
 			soumyajit@iitkgp.ac.in}
 	}
	
	\maketitle
	
	\begin{abstract}
		Increased dependence on networked, software based control has  escalated the vulnerabilities of Cyber Physical Systems (CPSs). Detection and monitoring components developed leveraging  dynamical systems theory are often employed as lightweight security measures for protecting such safety critical CPSs against false data injection attacks. However, existing approaches do not correlate  attack scenarios with parameters of detection systems. In the present work, we propose a Reinforcement Learning (RL) based framework which adaptively sets the parameters of such detectors based on experience learned from attack scenarios,  maximizing detection rate and minimizing  false alarms in the process while attempting performance preserving control actions.
	\end{abstract}
	
	\begin{IEEEkeywords}
		Cyber-physical systems, security, adaptive threshold, reinforcement learning, formal methods
	\end{IEEEkeywords}
	
	\section{Introduction}
	\label{secIntro}
\begin{comment}
Software intensive, network-centric Cyber-Physical Systems (CPS) are widely used in application domains ranging from industrial control, connected-mobility to defense installations. Most CPSs are safety critical in their operations. For example, most modern vehicles implement   a collection of safety critical CPSs (like, anti-lock braking system, electronic stability program, roll stability controller, etc.) that exchange sense and actuation messages using  light-weight intra-vehicular network protocols. An adversary can infiltrate such networks and degrade performance of such sub-systems to the point that may even be catastrophic for vehicle users as well as surrounding traffic.
\end{comment}
	\par While security is of paramount importance in CPS \cite{humayed2017cyber}, the real time hard deadlines of the safety critical functions and availability of limited computing resources constitute the major constraints to secure CPS design. Traditional heavy-weight cryptographic encryption techniques  (like  RSA,  AES)  along  with  MACs cannot be used in most of the cases as they lead to both computation and communication overhead \cite{munir2018design}. 
	\begin{comment}
	Therefore, it is difficult to secure every communication between the plant and controller. 
	\end{comment}
	One alternative is to design light-weight attack detectors by exploiting control-theoretic properties. Such detectors leverage the features of various state-observers like Luenberger or Kalman filter that the controller unit of almost every CPS comes with.  These observers estimate the state of the plant by computing residue as the difference between actual and observed sensor data. The detector compares this residue with a predefined threshold to identify an anomaly \cite{mo2010false}. Though observers can be useful in monitoring the system’s behavior, they cannot always be adjusted to detect unmodeled disturbances, like faults and attacks. Reducing observer gain may render it  insensitive towards small changes in system states, while increased  observer  gain will lead to increased estimation error covariance, which in turn will  degrade the system’s control performance.	
\begin{comment}
	Such detectors use the features of various state-observers like Luenberger or Kalman filter that the controller unit of almost every CPS comes with. From the received sensor data, these observers estimate the states of the plant, and a residue is computed as the difference between the estimated and actual plant output. This residue is then compared with a predefined constant threshold to detect an attack \cite{liu2011false, sandberg2010security, mo2010false}. Gains of these observers are designed to mitigate the effect of noise, that follows a certain distribution, in the system. While computing the gains, recursive methods are executed until they reach a steady state i.e. a constant value. In general, this steady-state gain is used in online applications. Therefore, though these observers can be useful in monitoring the system’s behavior, they cannot always be adjusted to detect unmodeled disturbances, like faults and attacks. While reduced observer gain may cause it to be insensitive towards small changes in system states, larger observer gain will increase estimation error covariance, thus degrading the system’s performance. Due to this limitation, statistical change detection methods, like $\chi^2$-test, Cumulative Sum (CUSUM), Sequential Probability Ratio (SPRT)  have  been used to detect any unmodeled  disturbances in CPS since past few decades \cite{willsky1975two}. These tests are applied to the residue and the outcome is compared with a fixed threshold. Among this, $\chi^2$-test is widely used because of simpler implementation.
\end{comment}
To overcome this limitation, the statistical change detection methods, like $\chi^2$-test, Cumulative Sum (CUSUM) \cite{willsky1975two,giraldo2018survey},  etc. are applied on the residue, before it is compared to the threshold. However, the constant threshold used by these residue based detectors may increase the false alarm rates (FAR) by  misinterpreting measurement noise as attack, leading to unnecessary  degradation in control performance. %Consider for example an autonomous vehicle (AV) with a detector that signals an attack scenario by raising frequent false alarms which trigger   safety measures that slow down the vehicle or even halt it, thus hampering the traffic flow. 
Moreover, recent research  \cite{teixeira2015secure, koley2020formal} have shown how a \emph{stealthy} attacker can fool such detectors by crafting perturbation sequences which create residues that are small enough (i.e. below threshold) to be classified as noise. %By stealthy we mean that the attacker will make the system unsafe by fooling the detector. 
Therefore, the fundamental question that arises is, \emph{whether the detection threshold for such monitors in CPS implementations can be dynamically adjusted  based on the deployment environment so that FAR is minimized while even small attack efforts can be detected}. In the present work we  propose an intelligent detection scheme that strives to achieve this goal. 

\par Some recent research  efforts, for example \cite{koley2020formal, Baek2018, Ghafouri2016}, have addressed this problem of balancing between FAR and detectability in the CPS context. However, they have the following limitations that we have addressed in the current work. Unlike \cite{Baek2018}, the proposed detector focuses on identifying attacks on CPS rather than faults. In \cite{Baek2018}, the authors have formulated a non-linear programming problem to synthesize adaptive thresholds, considering operating regions. When the system is online, a threshold is selected from the pre-calculated ones to detect transient faults, based on which operating region the system is currently working in. In case of attack, things are more difficult as an attacker can be smart enough to modify its action to remain stealthy. In \cite{koley2020formal}, the authors have presented two greedy algorithms based on formal methods to synthesize monotonically decreasing variable threshold based detectors to thwart targeted performance degrading attacks while minimizing FAR. However, they have defined the CPS system requirement in terms of settling time. %Therefore, the detector may not guarantee system  safety throughout the system progression.
In such works it is  often straightforward to  design attack vectors through constraint solving such that the settling time property is satisfied while some other safety property gets violated causing critical damage to the system.  %For example, consider a CPS where the controller reaches its set point within $50$ time units when an input event occurs (i.e. set point change, activation of higher level control loop etc). An attack can easily be designed that satisfies this settling time property, however can do enough damage much before the $50^{th}$ timestamp. 
On the contrary, our approach interprets system safety in terms of a safe operating region. Considering that an attacker may force the system to migrate beyond this safety boundary at any time instant, our proposed detector would try to detect such attack efforts as early as possible. 

CPSs are usually designed as closed loop feedback systems with both controller and observer in place, both working together to ensure stability while minimizing the effect of noise. Aim of an attacker would be to remain stealthy and destabilize the closed loop dynamics by failing the efficacy of the controller and the estimator. Therefore, while designing a detector we must also consider such an attack model. Moreover, in the view of real time constraint of the safety critical CPSs, the adaptive threshold generation process must be efficient in terms of timing overhead. Though, the authors of \cite{Ghafouri2016} have drawn similar motivation like ours and presented an attacker-defender game to solve the adaptive threshold selection problem, their work lacks these considerations.
\par Another important aspect while designing an intelligent secure CPS is the mitigation of an attack's effect, i.e. \emph{when an attacker is detected by the detector system and  the system is still within the safety boundary, how can its effect be mitigated at the  earliest ?} The approach in \cite{ferdowsi2018robust} proposes a Reinforcement Learning (RL) based robust control strategy for autonomous vehicle (AV) control in the presence of an attacker who modifies the spacing information between vehicles. The method leverages the fact that  measurement information (velocity of other vehicles) is drawn from multiple sensors and learns the  optimal weights of these sensors that mitigate the attack's effect. However, this work is applicable to a specific CPS and also the authors do not consider any  active security primitive like detection systems. In \cite{Zhou2019}, the authors formulate a secure state estimation problem followed by RL based optimal controller design to void the effects of detected attacks. Though the approach is based on a general CPS model, they have used a constant threshold based attack detector which may suffer from high FAR. In general, the fact that a well-trained RL agent is realizable for safety critical CPSs with real time requirements has already been established in other contexts like  energy efficiency\cite{wang2020energy}. 
	\begin{comment}
	\sd{In \cite{ferdowsi2018robust} also, authors have attended this challenge for an autonomous vehicle by designing a robust controller using Reinforcement Learning (RL). However, the proposed method is not general for all safety critical CPSs. RL has also been adopted by \cite{Zhou2019} to design optimal controller to void the effects of the attacks. Though their approach is very well established and based on a general CPS model, they have used a constant threshold based detector which may suffer from FAR.}
	\end{comment}
\par In this work, we consider that the communication channel between the plant and  controller is vulnerable, i.e. an attacker can gain access to the network and add spurious data to every communication between  plant and controller. Such an attack is called \emph{false data injection} (FDI) attack. Assuming that the attacker has complete knowledge of the system and associated detector, it can intelligently craft an attack to induce maximum damage to the system while remaining stealthy. 
\begin{figure}[H]
	    \centering
	    \includegraphics[width = \columnwidth]{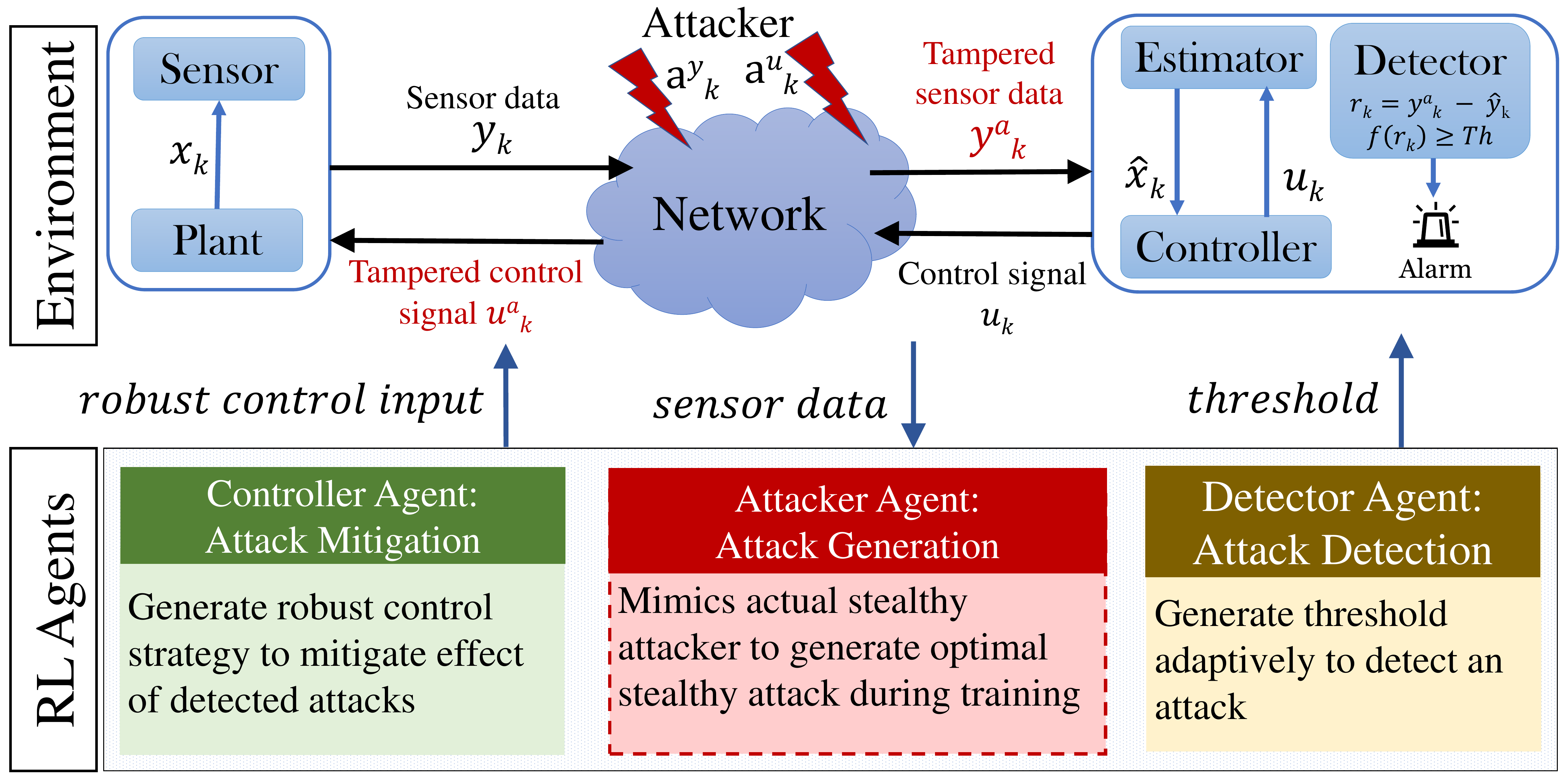}
	    \caption{Reinforcement learning based adaptive monitoring strategy}
	    \label{figPropsedSystem}
\end{figure}
We present an intelligent secure CPS model that consists of: $i)$ an adaptive attack monitor that thwarts an FDI attack, and $ii)$ a robust controller that mitigates the effect of such attacks. Considering an attack as unmodeled disturbance that does not follow any defined distribution; %like \cite{ferdowsi2018robust,Zhou2019}, 
we propose an RL based baseline framework as shown in Fig.~\ref{figPropsedSystem} that leverages the following RL agents.  
\begin{compactenum}
\item We present a novel detector for CPS attacks which leverages RL based adaptive threshold selection (Fig.~\ref{figPropsedSystem}). %Based on the current state of the system, our solution decides upon a threshold suitable for detecting a well crafted \sd{stealthy ?} FDI attack as early as possible. 
The choice of the threshold is based on  minimization of FAR while keeping it below a predefined upper bound.  
\item We present an RL based robust control strategy (Fig.~\ref{figPropsedSystem}) that would strive for preserving control performance in the presence of FDI  attacks. %\st{ Under such a controller, the system always remains within its preferable operating region. However, in an  attack scenario, the system may go beyond that region. Using the current threshold }
%\ik{
When no attack takes place, the system operates with an optimal controller that ensures high performance by restricting the trajectory inside a preferable operating region. %However, an attacker would always try to make the system unsafe, and thereby, optimal performance may not be guaranteed.}  
When an attack effort is detected  with the system  still within its safety boundary,  the proposed robust controller brings the system back to its preferable operating region. %The robust  controller, though  not  optimal in terms of control cost, ensures safety by mitigating the attack's effect.
\item For the RL based detector and controller to learn from experience, one needs to provide FDI attack vectors that are both stealthy and able to steer the system away from safe operations.  Our third RL agent mimics such attack behavior in the training phase of the system. 
%\ik{To ensure optimal performance of the proposed detector and robust controller when the system becomes online, we must train them against optimal FDI attack attempts. Therefore, we also design an RL agent that would mimic the behavior of an optimal stealthy FDI attack.} \st{We also design an RL agent for mimicing the attacker's behavior during the training phase. The objective of this RL agent will be synthesizing  optimal stealthy FDI attacks that can bypass the current threshold and try to steer the system out of the safety boundary.} \sd{relation with current work not clear, why a contribution}
\item We establish the 	usefulness of our multi-agent secure CPS model by considering attack  scenarios for well known CPS benchmarks and achieving  significant performance improvement w.r.t. baseline.
\end{compactenum}
	
\section{Secure CPS Model}
\label{secSecureCPSModel}
% 	\begin{figure}
% 		\centering
% 		\includegraphics[scale=0.25]{fig/controlStructure.pdf}
% 		\caption{Secure CPS architecture}
% 		\label{figSecureCPS}
% 	\end{figure}
	%General architecture of a secure CPS is presented in Fig.~\ref{figPropsedSystem}. The plant and the controller communicate between themselves over a network, which we consider is vulnerable to FDI attacks.
In the absence of an adversary, the closed loop dynamics of a CPS can be presented as a discrete linear time-invariant (LTI) system,
\begin{align}
% 		\begin{split}
\nonumber
x_{k+1} &= Ax_k + Bu_k + w_k,\ y_k = Cx_k + Du_k + v_k, u_k = -K\hat{x}_k,\\
\hat{x}_{k+1} &= A\hat{x}_k + Bu_k + Lr_k, r_k = y_k - C\hat{x}_k+v_k,\ \ 
% 		\end{split}
		\label{eqLTINoAttack}
\end{align}
where, $x_k \in \mathbb{R}^n$ is the system state vector,   $y_k \in \mathbb{R}^m$ is the measurement vector obtained from available sensors at $k$-th time stamp; $A, B, C, D$ are the system matrices. We consider that the initial state $x_0 \in \mathcal{N}(\bar{x}_0, \Sigma)$, the process noise $w_{k}\in \mathbb{R}^n \sim \mathcal{N}(0, \Sigma_w)$ and the measurement noise $v_{k} \in \mathbb{R}^m \sim \mathcal{N}(0, \Sigma_v)$ are independent Gaussian random variables. 
% 	All states of a plant may not be measurable i.e. $m$ and $n$ are not necessarily the same. Therefore, we also consider a state estimator unit to estimate the states of the plant $\hat{x}_k$.
Further, in every $k$-th sampling instant, the observable system state $\hat{x}_k$ is estimated using system output $y_k$ while minimizing the effect of noise, and used for computing the control input $u_k \in \mathbb{R}^l$.
% 	( refer Eq.~\ref{eqLTINoAttack}).
% 	\begin{equation}
% 		\begin{split}
% 			r_k = y_k - C\hat{x}_k+v_k;\ \hat{x}_{k+1} = A\hat{x}_k + Bu_k + Lr_k,\ 
% 			u_k = -K\hat{x}_k
% 		\end{split}
% 		\label{eqLTINoAttack}
% 	\end{equation}
The symbol $r_k$ denotes the residue i.e. the difference  between the measured and the estimated outputs. The observer gain $L$ and controller gain $K$ ensures that both $(A-LC)$ and $(A-BK)$ are stable. The system has a detector unit (Fig.~\ref{figPropsedSystem}) which computes a function $f(r_k)$ and compares it with  a threshold $Th$ to identify any anomalous behavior of the system. Considering an FDI attack, where the attacker injects false data $a^y_k$ and $ a^u_k$ (Fig.~\ref{figPropsedSystem}) to the sensor data and control signal respectively, %the same detector can be used to identify an attacker as well. In such case, 
the equation of the system dynamics will become,
	\begin{equation}
		\begin{split}
			x^a_{k+1} = Ax^a_k + B\tilde{u}^a_k + w_k;&\ y^a_k = Cx^a_k + D\tilde{u}^a_k + v_k + a^y_k;\\
			r^a_k = y^a_k - C\hat{x}^a_k+v_k;&\ \hat{x}^a_{k+1} = A\hat{x}^a_k + Bu^a_k + Lr^a_k\\
			u^a_k = -K\hat{x}^a_k;&\ \tilde{u}^a_k = u^a_k + a^u_k;
		\end{split}
		\label{eqLTIUnderAttack}
	\end{equation}
	Here, $x^a_k$, $\hat{x}^a_k$, $y^a_k$, $r^a_k$, $u^a_k$, $\tilde{u}^a_k$ represent plant state, estimated plant state, forged sensor data, residue, control signal, and forged control signal respectively in an attack scenario. In the present work we consider $f$ as the popular $\chi^2$-test commonly employed in existing works  on secure CPS \cite{mo2010false}. %which is applied to residue $r_k$ can be a norm  function or any statistical function, \sd{like $\chi^2$-test which is quite common in existing works \cite{}.} In this work, we consider the later one.  
	\par\noindent\textbf{$\chi^2$-Test and $\chi^2$-Distribution :}
	%\label{subsecChiSquare}
	The $\chi^2$-test is one of the most widely used statistical tests for examining the independence of two or more categorical variables. Given the observed count $O_i$ and the expected count $E_i$ of each category $i$, $\chi^2$ statistics can be computed as $\chi^2 = \sum_{i=1}^{k} \frac{(O_i - E_i)^2}{E_i}$.
	%\begin{equation}
	%	\chi^2 = \sum_{i=1}^{k} \frac{(O_i - E_i)^2}{E_i}
	%	\label{eqChiSquareStat}
	%\end{equation}
	Here, $k$ is the number of categories. Smaller value of $\chi^2$ signifies more correlation between the categories. Now, consider $n = k -1$ random variables that follow standard Gaussian distribution. Sum of the squares of these random variables follow a $\chi^2$ distribution of $n$ degrees of freedom (dof)  defined as, $P(x) = \frac{x^{\frac{n}{2}-1}e^{-\frac{x}{2}}}{2^{\frac{k}{2}\Gamma(\frac{k}{2})}}$. 
%		\begin{equation}
	%	P(x) = \frac{x^{\frac{n}{2}-1}e^{-\frac{x}{2}}}{2^{\frac{k}{2}\Gamma(\frac{k}{2})}}
	%	\label{eqChiDistribution}
	%\end{equation}
Here, $\Gamma$ denotes the Gamma function and mean of this distribution is $n$. Given the $\chi^2$-distribution of $n$ dof and the $\chi^2$ statistics, we can decide whether to accept or reject a specified null hypothesis. For example, a null hypothesis can be whether a random Gaussian vector has the expected mean and variance. Therefore, $\chi^2$-test can be used to detect anomalies in dynamical systems  \cite{da1994failure}, like CPSs\cite{mo2010false}. $\chi^2$-Test on an $m$ dimensional random variable $z(i) \sim \mathcal{N}(0, V)$ gives $\chi^2_z(i) = \sum_{j=i-l+1}^{i} z(i)^TV^{-1}z(i)$.
% 	\begin{align*}
% 	    \chi^2_z(i) = \sum_{j=i-l+1}^{i} z(i)^TV^{-1}z(i)
% 	\end{align*}
	$\chi^2_z(i)$ follows a $\chi^2$-distribution with dof = $ml$. With respect to a given threshold $\theta$, we say an anomaly is detected if $\chi^2_z(i) \geq\theta$ at some time stamp $i$.

\section{Proposed Methodology}
	\label{secProposedMethod}
In this section, we elaborately discuss the three principal components of our proposed adaptive secure CPS model: i) an adaptive threshold synthesis method, ii) an intelligent attack generation method, and iii) a robust control strategy.  Finally, we present a multi-agent RL framework that binds the above three components to ensure intelligent attack detection and mitigation.

\subsection{Optimal Threshold Synthesis}\label{subsecVariableThreshold}
We present a residue based attack detection system where we consider the Kalman filter as the estimator. The  proposed detector will adaptively select a threshold $Th_k$ at every $k$-th sample.
% 	For the discrete LTI system of Eqs.~\ref{eqLTINoAttack},~\ref{eqLTINoAttack}, the residues of the Kalman filter $r_k$ (Eq.~\ref{eqLTINoAttack}) follows a Gaussian distribution with mean $0$ and covariance $\Sigma_r$. 
Let, for the discrete LTI system shown earlier, %in Eqs.~\ref{eqLTINoAttack} (under no attack) and~\ref{eqLTIUnderAttack} (under FDI attack), 
the estimation error $e_k$ be defined as  $e_k = (x_k - \hat{x}_k)$. 
% 	($E(\hat{x})\rightarrow E(x),\ as\ k\rightarrow \infty$)
The Gaussian assumptions of noise and initial states ensure that $e_k$ follows a normal distribution with $0$ mean. We denote the steady state covariance matrix of this estimation error with $\Sigma_e$. So the system residue $r_k$, calculated in the Kalman estimator can be expressed as $r_k = Ce_k + v_k$ (see Eq.~\ref{eqLTINoAttack}). Given that both estimation error and measurement noise are Gaussian distributed  with zero mean and are independent of each other, the residue is normally distributed with $0$ mean and  covariance matrix given $ \Sigma_r = E[r_kr_k^T] - E[r_k]E[r_k]^T = E[(Ce_k)(Ce_k)^T] + E[v_kv_k^T] = C\Sigma_e C^T + \Sigma_v$. 
	
\begin{wrapfigure}{l}{0.6\columnwidth}
%		\small
		\centering
		\includegraphics[width=0.6\columnwidth,trim=1 5 5 5,clip]{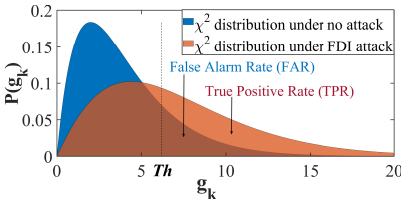}
		\caption{$\chi^2$-Distribution}
		\label{figChiSquareDist}
		\end{wrapfigure}

We use $\chi^2$-test on $r_k$ to find out how much the distribution of actual plant state $x_k$ and its estimate $\hat{x}_k$ vary from each other. Let $g_k$ denote the  $\chi^2$-test result at $k$-th sample and $g_k =  \sum_{i=k-l+1}^{k} r_i^T\Sigma_r^{-1}r_i$. Here, $l$ is the window size of $\chi^2$-test. In this case, the degree of freedom is $ml$, where $m$ is the number of available sensors in the plant. In a normal scenario (no attack), 
$g_k$ follows $\chi^2$ distribution with mean $ml$ (Fig.~\ref{figChiSquareDist}). Let $Th_k$ be the threshold that is currently (at $k$-th sampling instance) being used by our variable threshold based detector unit. Then, $g_k$'s probability density function (PDF) along with its cumulative distribution function with respect to $Th_k$ can be defined as, 
% 	\begin{align}
		$P(g_k) = \frac{g_k^{\frac{ml}{2}-1}e^{-\frac{g_k}{2}}}{2^{\frac{ml}{2}}\Gamma(\frac{ml}{2})};\ P(g_k \leq Th_k) = \frac{\gamma(\frac{ml}{2}, \frac{Th_k}{2})}{\Gamma(\frac{ml}{2})}$.
% 		\textit{, $\Gamma$ is the ordinary Gamma function}		
% \label{eqChiPDFNoAttack}
% 	\end{align}
Here, $\Gamma$ and $\gamma$ are ordinary and lower incomplete gamma functions respectively. %\st{Let $Th_k$ be the threshold that is currently (at $k$-th sampling instance) being used by our variable threshold based detector unit. The detector raises an alarm whenever $g_k$ exceeds $Th_k$.} 
We say it is a \emph{false alarm} when $g_k \geq Th_k$ even in the absence of an attacker.
 %(i.e. alarms, even in the absence of anomalous data). 
So \emph{false alarm rate} (FAR) is the ratio of the number of times when the alarm has been raised falsely and the total number of alarms raised. In Fig.~\ref{figChiSquareDist}, the blue curve and the maroon curve represent the  distribution of $g_k$ under no attack and attack respectively. Therefore, FAR should be the fraction of area under the probability distribution curve of un-attacked $g_k$ that is contained by the part beyond $g_k = Th_k$ and computed as $1 - P(g_k \leq Th_k)$. 
 
\par Now, in presence of an FDI attack, where spurious data $a^y_k$ and $a^u_k$ are added to the sensor and the actuator data respectively, this leads to non-centrality of the $\chi^2$-test (the maroon curve in Fig.~\ref{figChiSquareDist}) result 
$g_k^a$ obtained for the  residue $r_k^a$ under attack as established next. 
% 	This leads to a non-central $\chi^2$-distribution of $g_k^a$ which is the $\chi^2$ statistics of $r_k^a$. 
	%In the following theorem we show that, $g_k^a$ being the $\chi^2$-test result of $r_k^a$,
% 	which is the $\chi^2$ statistics of $r_k^a$ non-central $\chi^2$- distribution 
	%has a non-central $\chi^2$-distribution (the maroon curve in Fig.~\ref{figChiSquareDist}) 
% 	i.e. it follows a non-central $\chi^2$-distribution. 
% 	And that will lead to an increase in the detectability of such FDI attacks.
	\begin{theorem}
		Considering an FDI attack on an LTI system as specified in Eq.~\ref{eqLTIUnderAttack}, the $\chi^2$-test on residue $r_k^a$ follows a non-central $\chi^2$-distribution.
		\label{thNonCentric}
	\end{theorem}
	\emph{Proof:} Under FDI attack, the estimation error at $k$-th sample is given by $e_k^a = x_k^a - \hat{x}_k^a$. Let $\triangle e_k = e^a_k - e_k$ be the difference between the estimation error under attack and no attack scenarios. Thus, $r^a_k = Ce^a_k + v_k + a^y_k = Ce_k + C\triangle e_k + v_k + a^y_k$. 
% 	then 
	%\begin{align}
% 		\begin{split}
	%		r^a_k = Ce^a_k + v_k + a^y_k = Ce_k + C\triangle e_k + v_k + a^y_k
% 		\end{split}
	%	\label{eqResidueInTermsOfDeltaE}
	%\end{align}
Considering that the mean of the estimation error is $0$ and measurement noise is independent of the estimation error and sensor attack, the covariance $\Sigma_{r^a}$ of the residue $r^a_k$ generated due to an FDI attack can be computed as $\Sigma_{r^a} = E[r_k^ar_k^{a^T}] - E[r_k^a]E[r_k^a]^T$ where, \begin{align*}
		\begin{split}
			E[r_k^a&r_k^{a^T}] = E[Ce_ke_k^TC^T + Ce_k\triangle e_k^TC^T + C\triangle e_ke_k^TC^T +\\
			& C\triangle e_k\triangle e_k^TC^T + C\triangle e_ka^{y^T}_k + v_kv_k^T + a^y_k\triangle e_k^TC^T + a^y_ka^{y^T}_k]\\
			E[r_k^a]&E[r_k^a]^T =(CE[\triangle e_k] + E[a^y_k])(E[\triangle e_k]^TC^T + E[a^y_k]^T)\\
			&= C\mu_{\triangle e}\mu_{\triangle e}^TC^T + C\mu_{\triangle e}\mu_{a^y}^TC^T + \mu_{a^y}\mu_{\triangle e}^TC^T + \mu_{a^y}\mu_{a^y}^TC^T			
		\end{split}
	\end{align*}
Here, the notations $\mu_i$ and $\Sigma_i$ denote the mean and variance respectively of any variable $i$, and $\Sigma_{i,j}$ denotes the covariance of $i,\ j$ for any $i,j$. Using the expressions of $E[r_k^ar_k^{a^T}]$ and $E[r_k^a]E[r_k^a]^T$ we get,
\begin{comment}
$
\Sigma_{r^a} = C\Sigma_eC^T + C\Sigma_{e,\triangle e^T}C^T + C\Sigma_{e,a^{y^T}} + C\Sigma_{\triangle e, e^T} + C\Sigma_{\triangle e}C^T + C\Sigma_{\triangle e,a^y} + \Sigma_{a^y,e^T} + \Sigma_{a^y,\triangle e}C^T + \Sigma_v + \Sigma_{a^y} = \Sigma_r + P
$	
\end{comment}
	\begin{align}
		\begin{split}
			\Sigma_{r^a} &= C\Sigma_eC^T + C\Sigma_{e,\triangle e^T}C^T + C\Sigma_{e,a^{y^T}} + C\Sigma_{\triangle e, e^T} \\
			&+ C\Sigma_{\triangle e}C^T + C\Sigma_{\triangle e,a^y} + \Sigma_{a^y,e^T} + \Sigma_{a^y,\triangle e}C^T\\
			&+ \Sigma_v + \Sigma_{a^y} = \Sigma_r + P
		\end{split}
	\label{eqCovOfResUnderAttack}
	\end{align}
Here, $P = C\Sigma_{e,\triangle e^T}C^T + C\Sigma_{e,a^{y^T}} + C\Sigma_{\triangle e, e^T} + C\Sigma_{\triangle e}C^T	+ C\Sigma_{\triangle e,a^y} + \Sigma_{a^y,e^T} + \Sigma_{a^y,\triangle e}C^T + \Sigma_{a^y}$, $\Sigma_r = C\Sigma_eC^T + \Sigma_v$.  % (Eq.~\ref{eqResidueCovar}). 
 Since, by definition, covariance is positive semi definite and variance is positive, %  While computing $\Sigma_r$ and $\Sigma_{r^a}$, we have ignored the covariance of independent terms. Therefore, 
 both $\Sigma_r$, $P$ and consequently $\Sigma_{r^a}$ are positive definite. Considering an FDI attack on the both control signal and sensor output, we apply $\chi^2$-test on $r_k^a$ which gives $g_k^a = \sum_{i=k-l+1}^{k}  r_i^{a^T}\Sigma_r^{-1}r_i^a$. Therefore, the mean $\mu$ of the $\chi^2$ statistics $g_k^a$ of $r_k^a$ over an observation window of length $l$ can be computed as $\mu = E[g_k^a] = E[\sum_{i=k-l+1}^{k} r_i^{a^T}\Sigma_r^{-1}r_i^a] =  \sum_{i=k-l+1}^{k}trace[ \Sigma_{r^a}\times \Sigma_r^{-1}]$ i.e.,
	\begin{align}
		\begin{split}
			\mu &=\sum_{i=k-l+1}^{k}trace(\Sigma_r \Sigma_r^{-1}) + \sum_{i=k-l+1}^{k}trace( P\Sigma_r^{-1})\\
			& = ml + \sum_{i=k-l+1}^{k}trace( P\Sigma_r^{-1}) > ml
		\end{split}
		\label{eqChiMeanUnderAttack}
	\end{align}
\begin{comment}
$	\mu = E[g_k^a] = E[\sum_{i=k-l+1}^{k} r_i^{a^T}\Sigma_r^{-1}r_i^a] = \sum_{i=k-l+1}^{k}trace[ \Sigma_{r^a}\times \Sigma_r^{-1}]   =\sum_{i=k-l+1}^{k}trace(\Sigma_r \Sigma_r^{-1}) + \sum_{i=k-l+1}^{k}trace( P\Sigma_r^{-1}) = ml + \sum_{i=k-l+1}^{k}trace( P\Sigma_r^{-1}) > ml
$ 		 
\end{comment}
The last inequality  follows from the fact that $P$ is positive definite. Hence, the mean of $g_k^a$ is strictly greater than the mean $ml$ of $g_k$. This makes the distribution of $g_k^a$ a non-central one with non-centrality parameter $\lambda = \sum_{i=k-l+1}^{k}trace( P\Sigma_r^{-1})$.\hfill$\Box$
	\begin{theorem}
		Leveraging the non-central distribution of $\chi^2$ statistics increases the detectability of a false data injection attack.
		\label{thIncreaseDetect}
	\end{theorem}
	\emph{Proof:} The non-central distribution of $g_k^a$ with mean $\mu$ and non-centrality parameter $\lambda$ can be defined in terms of the following PDF\cite{siegel1979noncentral}, 
\begin{comment}
	\begin{equation}
		P(g_k^a) = \frac{1}{2}e^{-\frac{(g^a_k+\lambda)}{2}}(\frac{g^a_k}{\lambda})^{\mu /4-1/2}I_{\mu /2-1}(\sqrt{\lambda g_k^a}).
		\label{eqChiPDFAttack}
	\end{equation}
\end{comment}
$P(g_k^a) = \frac{1}{2}e^{-\frac{(g^a_k+\lambda)}{2}}(\frac{g^a_k}{\lambda})^{\mu /4-1/2}I_{\mu /2-1}(\sqrt{\lambda g_k^a})$ with $I$ denoting Bessel function. With respect to $Th_k$, we say an FDI attack is detected if $g_k^a > Th_k$. This is a true positive case. Therefore, the true positive rate (TPR) of detecting an attack is computed as $TPR = 1-P(g_k^a \leq Th)$ where 
%	can be defined as the fraction of area under the distribution curve (Fig.~\ref{figChiSquareDist}) of $g_k^a$ beyond $g_k^a = Th_k$. We compute TPR as $1-P(g_k^a \leq Th)$ where:	
%	\begin{equation}
%		P(g_k^a \leq Th_k) = 1 - Q_{\mu/2}(\sqrt{\lambda},Th_k)
%		\label{eqCumDistrAttack}
%	\end{equation}
$P(g_k^a \leq Th_k) = 1 - Q_{\mu/2}(\sqrt{\lambda},Th_k)$. Essentially this is fraction of area under the distribution curve (Fig.~\ref{figChiSquareDist}) of $g_k^a$ beyond $g_k^a = Th_k$. Here, $Q$ is  Marcum Q-function \cite{siegel1979noncentral}. In Theorem~\ref{thNonCentric}, we have proved that $\mu > ml$ where $ml$ is the mean of $g_k$. This causes the non-central $\chi^2$ distribution of $g_k^a$ to be more shifted towards the right than the $\chi^2$ distribution of $g_k$. 
\par Moreover, the variance of $g_k$ is $\sigma = 2ml$ and variance of $g_k^a$ is $\sigma_a = 2(ml + 2\lambda)$, where $\lambda > 0$. Clearly, $\sigma_a > \sigma$. Therefore, the expected deviation of $g_k^a$ from $\mu$ is more than the expected deviation of $g_k$ from $ml$ which makes the distribution of  P($g_k^a$) wider and thereby flatter (since the area under both curves is unity). Hence, the fraction of area under the curve beyond $g_k^a = Th_k$ of P($g_k^a$) is more than that in case of P($g_k$) as shown in Fig.~\ref{figChiSquareDist}. So, the non-central $\chi^2$ distribution improves TPR i.e. attack detectability thus leading to $TPR > FAR$ for a properly chosen threshold parameter $Th$. \hfill$\Box$
	\par
	\begin{comment}
	Now, using Eq.~\ref{eqCumDistrNoAttack} and~\ref{eqCumDistrAttack}, we define FAR and TPR as,
	\begin{equation}
		FAR = 1 - P(g_k < Th); \ TPR = 1 - P(g_k^a < Th)
		\label{eqTPRFAR}
	\end{equation}
	\end{comment}
%	The length of the $\chi^2$-test window i.e. $l$ contributes to the mean of the $\chi^2$-distribution (both normal as well as non-central). Therefore, $l$ affects the distributions in Eqs.~\ref{eqChiPDFNoAttack},~\ref{eqChiPDFAttack}, which in turn influence FAR and TPR calculation given in Eq.~\ref{eqTPRFAR}. 
\noindent Given the dependence of $TPR, FAR$ on the parameters $l\, (\textrm{ window length}), Th$ 
% The objective of the adaptive variable threshold based detector would be selecting a suitable threshold $Th$ and $\chi^2$ window length $l$ such that FAR is minimized and attack detectability i.e. TPR is maximized. Hence, we present 
the problem of synthesizing an optimal detector can be formulated as following optimization problem:
	\begin{equation*}
		J_t = \max_{l, Th} \ w_1\times TPR - w_2\times FAR \ \textrm{s.t.} \ FAR<\epsilon,\ l<l_{max}
		\label{eqOptThreshold}
	\end{equation*}
aimed at minimizing $FAR$ and maximizing $TPR$. Here  $w_1, w_2 \in [0, 1]$ are the constant weights of TPR and FAR respectively, $\epsilon$ is the maximum allowable FAR, and $l_{max}$ is maximum allowed sequence length.  Given $y_k^a$, the current sensor measurement vector, the solution of the above optimization problem is a pair $<l^*, Th^*>$, where $l^*$ and $Th^*$ are the optimal $\chi^2$ window length and threshold respectively with respect to current measurement of the system states.
	
\subsection{Intelligent Attack Generation}
	\label{subsecAttackEstimator}
\begin{wrapfigure}{l}{0.4\columnwidth} 
%		\small
		\centering		\includegraphics[width=0.4\columnwidth ]{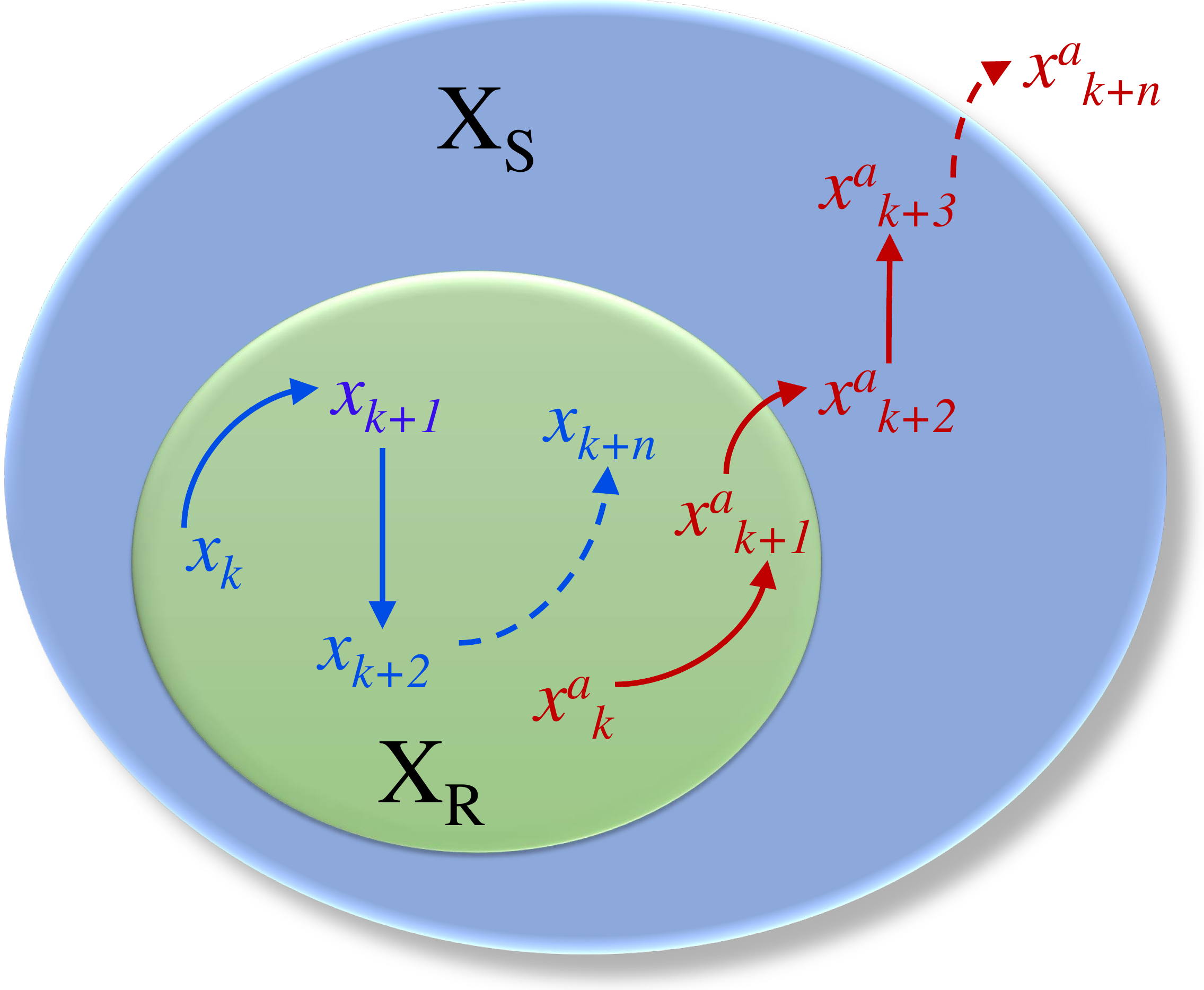}
		\caption{ Operating regions of the system: Safe region $X_S$, Preferable Operating Region $X_R$}
		\label{figRegions}
	\end{wrapfigure}	
Considering the discrete LTI system specified in Eq.~\ref{eqLTINoAttack}, we classify the operating region of the system, as demonstrated in Fig.~\ref{figRegions}, in two primary subregions: i) \emph{safe region} $X_S$, and ii) \emph{preferable operating region} $X_R$ where $X_R \subset X_S$. The system becomes unsafe when it goes beyond the outermost region $X_S$. The middle region $X_R$ defines the set of possible states in which system operation is preferred due to performance consideration. We formally  define such preferable operating regions as robust invariance sets as follows. 
\begin{definition}	\emph{Preferable Operating Region: } Considering the discrete LTI system specified in  Eq.~\ref{eqLTINoAttack} and controller with gain $K$, the preferable operating region of the system is defined as:		
\begin{center}
$X_R = \{x \ |\ \forall\ w \sim \mathcal{N}(\Sigma_w, 0), \ v\sim \mathcal{N}(\Sigma_v, 0), \ and \ x \in X_R, \ f^n(x, K) \in X_R \  \forall n \in \mathbb{N}\}$
\end{center}
where, $X_R\subset X_S$ and \emph{'f'} implements the state transition process of the LTI system (Eqs.~\ref{eqLTINoAttack}) in no attack scenario.
\hfill$\Box$
\end{definition}
By preferable operating region, we mean that starting from anywhere within $\in X_R$, the controller ensures that the system will always remain within $X_R$ in the absence of any FDI attack. Note the choice of  $X_S$ can be exercised  depending upon system description and safety criteria. The invariant set based preferable operating region $X_R$ is chosen in practice as some {\em $i$-step invariant set} within $X_S$ for which the controller guarantees satisfactory performance. %\ik{In this work, we design $K$ as a Linear-Quadratic-Gaussian (LQG) controller \cite{aastrom2013computer}. In general, any optimal control framework is applicable  though.} %Based on this idea, we now bring the notion of formal methods to present an approach to compute $X_R$.
\\
\textbf{Synthesis of Preferable Operating Region $X_R$:} 	\begin{comment}
		Given the state estimates $\hat{x}_k$, LQG controller computes the optimal control input $u_1, u_2, \cdots$ by solving the following cost function,
	\begin{equation}
		J = \min_{u_1, u_2, \cdots} E(\sum_{i = 1}^{\infty} [(x_i-x_{ref})Q(x_i-x_{ref})^T + u_iRu_i^T])
		\label{eqLqrController}
	\end{equation}
Here, $Q$ and $R$ are positive semi-definite matrices and $x_{ref}$ is the reference point of operation of the controller. This optimization problem returns a constant gain $K = -(R + B^TPB)^{-1}B^TPA$ for which $u_k = -K\hat{x}_k$ is optimal $\forall k$ and $P$ satisfies the Algebraic Riccati equation $P = Q + A^TPA - A^PB(R + B^TPB)^{-1}B^TPA$.
		\end{comment}
We present a Satisfiability Module Theory (SMT) aided constraint solving approach for computing $X_R$ for a given safety-critical CPS in Algo.~\ref{algPrefOptReg}. The \Call{getPerformanceRegion}{ } function takes as input the system matrices $A, B, C$, controller gain $K$, observer gain $L$, safety region $X_S$, and the forward step count $i$ for an $i$-step invariant set. The  preferable operating region $X_R$ is defined as a fraction of $X_S$ i.e. $X_R = depth\times X_S$, $depth \in (0,1)$. For an $n$ dimensional system  $X_S\in \mathcal{I}^n$ with $\mathcal{I}$ representing any {\em real} interval. Initially, we consider $depth = d_\delta$ where $d_\delta \in (0,1)$ and compute $X_R$ accordingly in line~\ref{algPrefOptRegXRInit}. Then, the plant state $x_0$ is initiated {\em symbolically} from $X_R$ (lines~\ref{algPrefOptRegXRInit}-\ref{algPrefOptRegInit}). We unroll the state progression $i$ times following Eq.~\ref{eqLTINoAttack} in lines \ref{algPrefOptRegStateprog}-\ref{algPrefOptRegStateprog3}. For $X_R$ to be the desired preferable region, after $i$ steps, plant state must reside within $X_R$ i.e., $x_i \in X_R$. This symbolic constraint is provided as an assertion $\phi$  (line~\ref{algPrefOptRegAssert}). Negation of this assertion, i.e. $\neg\phi$, is  passed to the SMT solver Z3 \cite{de2008z3}. Z3 tries to find a value of $x_0$ for which $\neg\phi$ can be satisfied. If such an assignment of $x_0$ is found, it implies that there exists an initial state of the system $\in$ the current candidate $X_R$, starting from which the system does not converge back to $X_R$ in $i$ steps. Note that we consider the reference point of the system is $0$. The optimal LQG controller with gain $K$ guarantees to keep the system close to the reference point at steady state. Therefore, the algorithm retries by increasing $depth$ by a step $d_\delta$ and looks for a larger $X_R$ (line~\ref{algDepthUpdate}-\ref{algPrefOptRegAssertCheck}). Otherwise, current $depth\times X_S$ is returned as final $X_R$ (line \ref{algPrefOptRegReturn}).
	\begin{algorithm}[!h]
%		\footnotesize
	\caption{Region Synthesis for Assured Performance}
	\label{algPrefOptReg}
	\begin{algorithmic}[1]
	\Require{Closed Loop System Matrices $\langle A,B,C,K,L\rangle$, system Safety Region $X_S$, forward step count $i$ in \emph{i-step invariant set} computation}
	\Ensure{Preferable Operating region $X_R$}
	\Function{getPerformanceRegion}{$\langle A,B,K,L\rangle$,$X_S$,$i$}
	\State $depth \gets d_{\delta}$; $X_R \gets depth\times X_S$; $x_0\in X_R;$ \label{algPrefOptRegXRInit}
	\State \label{algPrefOptRegInit}$ y_0\gets Cx_0; \hat{x}_0\gets 0; u_0\gets -K\hat x_0;r_0\gets y_0-C\hat{x}_0;$
	\Repeat
	\State $X_R\gets depth \times X_S$; \label{algPerfRegUpdate}
	\For{$k=1$ to $i$}
		\State \label{algPrefOptRegStateprog} $x_k \gets Ax_{k-1} + Bu_{k-1} + w_k$; 
		\State $\hat{x}_k \gets A\hat{x}_{k-1} + Bu_{k-1} + Lr_{k-1}$; \label{algEstimatedStateProg}
		\State\label{algPrefOptRegStateprog3}
		$u_k\gets -K\hat{x}_k;\ y_k\gets Cx_k + v_k;\ r_k\gets y_k-C\hat{x}_k;$
	\EndFor
    \State \label{algPrefOptRegAssert}$\Phi \gets$\textbf{assert}(($x_0\in X_R$) $\Rightarrow $ ($x_i\in X_R$));
    \State $depth \gets depth\ + \ d_{\delta}$; \label{algDepthUpdate}
	\Until{$\neg\Phi$ is {\em $unsatisfiable$} $\wedge\ depth \geq 1$} \label{algPrefOptRegAssertCheck}
    \State \Return \label{algPrefOptRegReturn}$X_R$
	\EndFunction
	\end{algorithmic}
\end{algorithm}
We set $d_\delta$ as $0.1$ in our experiments. Thus, it is formally guaranteed that the system will always remain within $X_R$ when no attack is taking place provided it has been initiated from $X_R$ itself. In this work, we design $K$ as a Linear-Quadratic-Gaussian (LQG) controller. In general, any optimal control framework is applicable  though.
\par In the CPS context, the attacker's motive is to steer the system  beyond the safe set $X_S$ while trying to remain stealthy by reducing the TPR. %Based on the current state of the system, we estimate the optimal false data $\triangle  y_k$ and $\triangle u_k$, \sd{diff symbol, plz make changes consistent every where} that the attacker can inject to the sensor and actuation signal respectively to achieve this goal. 
Given the sensor measurement $y_{k-1}$, we present this attack estimation problem as the following optimization problem:
\begin{center}
    $J_a = \max_{a^y_k, a^u_k}[ -w_1\times TPR + w_2\times FAR +\sum_{i=k}^{\infty}(|x_{i+1}|-|X_S|)^TW_3(|x_{i+1}|-|X_S|) ] \ s.t. \ y^a_i \in \epsilon_y,\ u^a_i \in \epsilon_u$
\end{center}
% 	\begin{equation*}
% 		\begin{split}
% 			&J_a = \max_{a^y_k, a^u_k}[ -w_1\times TPR + w_2\times FAR \\+ &\sum_{i=k}^{\infty}(|x_{i+1}|-|X_S|)^TW_3(|x_{i+1}|-|X_S|) ] %=[-J_t+\sum_{i=k}^{\infty}(|x_{i+1}|\\
% 			%&-|X_S|)^TW_3(|x_{i+1}|-|X_S|) ]
% 			\ s.t. \ y^a_i \in \epsilon_y,\ u^a_i \in \epsilon_u 
% 		\end{split}		
% % 		\label{eqOptAttack}
% 	\end{equation*}
Here, $w_1$ and $w_2$ are the same weights used in $J_t$ for the optimal threshold cost function $J_t$. Since the attack generation method will be used for experience learning of  threshold tuner and robust control RL agents, it is imperative that $J_a$ assumes knowledge about $J_t$ and tries to negate its cost objective. This is captured in the first two component terms of $J_a$. The last component of $J_a$ accounts for deviation of the current system state from the safety boundary $X_S$ using a quadratic weighted distance metric where $W_3$ is a diagonal matrix consisting of relative weights corresponding to criticality of each dimension. Also,  $\epsilon_y, \epsilon_u$ indicate the allowable sensor range, and actuation saturation range respectively. The solution $\langle a_k^{y^*},\,a_k^{u^*}\rangle$ of the above optimization problem is a possible attack vector that can breach the safety barrier while being stealthy, i.e. by nullifying the detector objective function $J_t$. In both $J_t,\, J_a$ implicit constraints are system and detector dynamics.
	
\subsection{Robust Controller Design}
\label{subsecAttackMitigation}
%A stealthy FDI attacker would try to steer the system beyond $X_S$, thereby beyond the preferable operating region $X_R$. The $X_R$ is designed with respect to the controller gain $K$. Under an FDI attack, if the system is found to be out of $X_R$, it implies that the gain $K$ is not enough to ensure system's preferable operation under such an attack. 
The LQR controller gain $K$ is designed to provide  optimal control action with respect to control cost under no attack scenario. However, it may not guarantee robustness against FDI attacks. To mitigate  effect of an FDI attack, we propose a robust control strategy that will be triggered only when the adaptive threshold based detector (Sec.~\ref{subsecVariableThreshold}) detects an FDI attack and the system $\in X_S\setminus X_R$ at the moment of attack detection. 
	\begin{comment}
	The objective of this controller would be bringing the system back to $X_R$ in minimum time. Such robust controller may not be optimal in terms of control cost, but ensures system's preferable performance under FDI attack.
	\end{comment} 
	Given the forged sensor data $y_k^a$, we compute such a robust control action by solving the following optimization problem:
	\begin{center}
	    $J_c = \min_{u_k^a} \ \sum_{i=k}^{\infty}(|\hat{x}^a_{i+1}|-|X_R|)^TW_3(|\hat{x}^a_{i+1}|-|X_R|)\ s.t. \ y^a_i \in \epsilon_y,\ u^a_i \in \epsilon_u$
	\end{center}
    % \begin{align}
    % \small
    %     \begin{split}
    %         J_c = \min_{u_k^a} \ \sum_{i=k}^{\infty}(|\hat{x}^a_{i+1}|-|X_R|)^TW_3(|\hat{x}^a_{i+1}|-|X_R|)\\
    %         s.t. \ y^a_i \in \epsilon_y,\ u^a_i \in \epsilon_u
    %     \end{split}
    %     \label{eqRobustController}
    % \end{align}
Naturally, system dynamics is an implicit constraint here. We have used the same weight matrix $W_3$ from $J_a$ to nullify the attack's effect  (assuming the knowledge of the attack generation module about all other cost functions). The solution of the above optimization problem is a control input $u_k^{a^*}$ that minimizes the damage induced by the attacker by bringing the system back inside $X_R$ (thereby, inside $X_S$). Note that $u_k$ is not optimal w.r.t. performance unlike an optimal controller; being robust it prioritizes safety.   Thus, we are allowing higher control effort as long as it does not exceed the actuation saturation limit  as the objective of this controller would be bringing the system back to $X_R$ in minimum time.  %Such robust controller may not be optimal in terms of control cost, but ensures system's preferable performance under FDI attack. The system switches between the optimal LQR controller and the robust controller based on which region it is in.
Once inside $X_R$, the system switches to the optimal controller. 
    
\begin{figure}
\centering
% \begin{subfigure}[b]{0.5\textwidth} 
\includegraphics[width=\columnwidth,scale=0.35,clip]{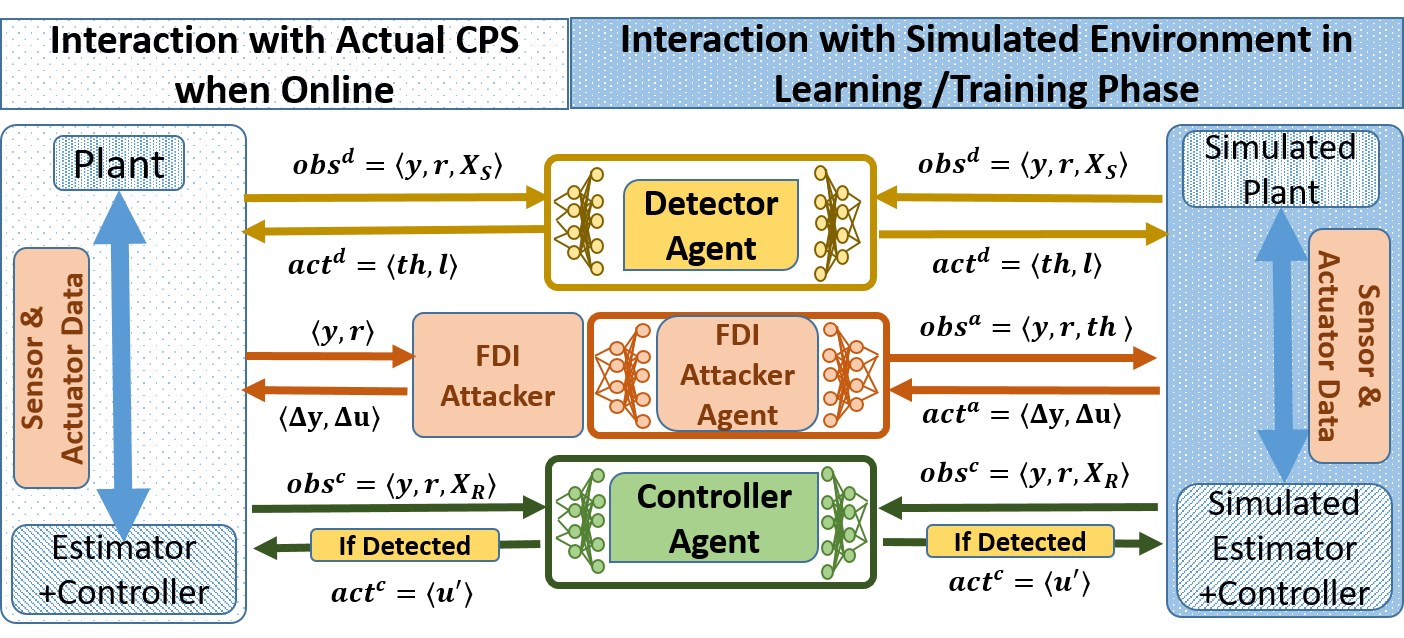}
\caption{RL Based Adaptive Monitoring Framework}
% \label{figRLAll}
% \end{subfigure}
% \hfill
% \begin{subfigure}[b]{0.48\textwidth} \includegraphics[width=\columnwidth,scale=0.3,clip]{fig/ddpg.png}
% \caption{A DDPG Agent}
% \label{figRLAgent}
% \end{subfigure}
% \caption{Proposed RL Based Framework in Action}
\label{figRL}
\end{figure}

\subsection{The Reinforcement Learning Framework} 
\label{subsecRL} Since an attacker may exhibit  unknown dynamics, the  central idea of the work is to learn an adaptive  attack detection scheme along with a robust  controller for attack mitigation whenever possible. %In previous sections we have established the  objectives to design the FDI attack detection strategy that promises minimum FAR and maximum attack detectability. It is equipped with a controller as well that refrains the system from becoming unsafe due to detected attacks. These objective functions are to be optimized in order to design an effective attack monitoring system. 
%Now there are a few challenges when we try to solve these optimization problems. \textbf{(i)} The false data injected into the sensors and actuators are hard to model because they are independent, random and does not follow any pattern. \textbf{(ii)} So, the parameters of this adaptive monitoring strategy eg. threshold, detection window will not be directly derivable from the injected false data signature.  \textbf{(iii)} Moreover, choices of the FDIs for an attacker to damage the system without being noticed is highly system specific. Whereas the change detection techniques in place are more likely to be generic for easier implementation purposes. 
% So we use a machine learning based  approach to model such unmodeled False data and learn how to utilize that intuition to increase the effectiveness of designed monitoring system against such attacks. Scarce of system specific labeled false data makes reinforcement learning (RL) an obvious choice because, it has the ability to take feedback from the environment and optimize its learning on the run. During this training period it learns and develops an optimal policy which is then put online to play its part.
Due to scarcity of system specific labeled false data and the requirement of learning parameters in dense domains, we employ the popular Deep Deterministic Policy Gradient (DDPG) algorithm \cite{lillicrap2019continuous} which outputs deterministic actions instead of optimized action distribution over the continuous action space. The overall RL framework is shown in  Fig. \ref{figRL}. A Deep Q-Network  to criticize and update the actor policy by calculating the Q value against the state and chosen action. %(see Algo.~\ref{algDDPG} and Fig.~\ref{figRLAgent}).
\par We first describe the \emph{environment} and \emph{agent} specifications to understand the learning process that helps derive the design parameters. Since a plant-controller closed loop  system, equipped with a $\chi^2$-based detector, as shown in Fig.~\ref{figPropsedSystem}) is the system under test here, we design our environment by modeling such a closed loop system. Our methodology uses three DDPG agents ($\Lambda$) that interact with this environment. By observing certain parameters from the environment ($obs$), the agents learn how to intelligently choose an action ($act$) to influence it as they want. The following table lists the observation and action variables.
\begin{table}[H]
    \centering
    \scalebox{0.9}{
    \begin{tabular}{|c|c|c|}
    \toprule
    \rowcolor[HTML]{C0C0C0} RL Agent & Observations & Actions\\ \hline
    Attacker Agent $\Lambda^a$ 
    & $obs^a= \langle y_k,u_k,Th_k, X_S \rangle$ & $act^a= \langle a^y_k a^u_k \rangle$ 
    % & $\mathcal{R}^a=\\ W_1 \mathcal{R}^a_1+W_2 \mathcal{R}^a_2 + W_3 \mathccal{R}^a_3$
    \\ \hline
    Detector Agent $\Lambda^d$ 
    & $obs^d=\langle r_k^a,y_k^a,X_S\rangle$ 
    & $act^d=\langle Th_k, l\rangle$ 
    % &$\mathcal{R}^d = -w_1*\mathcal{R}_d^{1}+w_2*\mathcal{R}_d^2$ 
    \\ \hline
    Controller Agent $\Lambda^c$ 
    & $obs^c=\langle y_k,X_{R},X_S\rangle$ 
    & $act^c=\langle u_k^a\rangle$ 
    % & $\mathcal{R}^c = w_3 \mc{R}^c_1$ 
    \\ \hline
    \end{tabular}
    }
    \caption{ RL Agent Details}
    \label{tabRLAgents}
\end{table}
\begin{algorithm}[!b]
%	\footnotesize
	\caption{RL Based Adaptive FDI Attack Monitoring  Framework}
	\label{algMultiAgentFramework}
	\begin{algorithmic}[1]
	\State $X_{R}\gets$ \Call{getPerformanceRegion}{system, safety region $X_{S}$, forward step count $i$ in \emph{i-step invariant set} computation} \Comment{Compute the preferable performance region for given closed loop system}
	\State $obs_{ini}\gets rand(obs\in Obs)$ \Comment{Initialize the environment/system with a random state from $X_R$}  
	\State $[\Lambda^a,\Lambda^c,\Lambda^d]\gets$ \Call{trainAgents}{system, [$\Lambda^a,\Lambda^c,\Lambda^d$],$training\ specs$} \Comment{Competitive and Collaborative offline training of the multi-agent setup}
	\State Put the system online
	\For{every sampling iteration $k\in[1, T]$}
	\State Collect the system observable states $[obs^c_k,obs^d_k]$ in current iteration
	\State $[act^d_k]\gets [\Lambda^d(obs^d_k)]$ \Comment{Update the system with actions from detector agent}
	\If{$act^d_{k-1}$  flags an FDI $\And$ system state $x_k\in X_S-X_{R}$}
	\State $[act^c_k]\gets [\Lambda^d(obs^d_k)]$ \Comment{Update the system with actions from controller agent}
	\EndIf
	\State \parbox[t]{223pt}{Simulate the system with current actions and generate next set of observable states $[obs^c_{k+1},obs^d_{k+1}]$}
	\EndFor
	\end{algorithmic}
\end{algorithm}
The \emph{Attacker Agent ($\Lambda^{a}$)} is  designed to intelligently inject false data into the system. 
% The agent chooses certain  amount of false data, bounded by the sensor and actuator design specifications (Refer Eq.~\ref{eqOptAttack}) as \emph{actions ($act^a=[\Delta y, \Delta u]$)}. The sensor and actuator data of the system along with the detector and the system safety specifications are the \emph{observations ($obs^a=[y,u,threshold,X_S]$)}. $\Lambda_a$ considers these observations every simulation instance, to learn about the system characteristics. Its aim is to make the system state unsafe by polluting the plant output and actuation signal, before they reach the control (estimation) unit and the plant respectively, without being flagged. The agent is \emph{rewarded} based on how optimally it can choose such an action observing the current states. Hence
We design a \emph{reward} function  $\mathcal{R}^a(obs^a_k,act^a_k,obs^a_{k+1})$ for $\Lambda_a$, that is built with the components of $J_a$. The FDI attacker agent tries to solve the optimization problem $J_a$ in every sampling iteration during training, by exploiting several actions while exploring the action space. These transitions are then stored as experiences. The training algorithm learns the highest expected return from the experiences and updates the RL policy to earn it. This helps it eventually choose the optimal action i.e. the optimal false data to inject into the sensors and actuators $\langle a^y_k,a^u_k \rangle$, that generates the maximum value of $J_a$ as the reward $\mc{R}^a$ for the FDI attacker agent. Similarly for \emph{Threshold-based Detector Agent} ($\Lambda^{d}$), a reward function  $\mc{R}^d(obs^d_k,act^d_k,obs^d_{k+1})$  is designed  with the components of $J_t$. The agent  intelligently chooses the optimal change detection parameters like threshold $Th$ and $\chi^2$ detection window length $l$ by optimizing the  objective function $J_t$ i.e. aiming the maximum reward $\mc{R}^d(obs^d_k,act^d_k,obs^d_{k+1})$. The \emph{reward} function $\mathcal{R}^c(obs^d_k,act^d_k,obs^d_{k+1}) 
% 	= -w_1*\mathcal{R}_d^{1}+w_2*\mathcal{R}_d^2
$ is also designed  with $J_c$ for $\Lambda^c$ to choose an optimal control input $u^a_k$ that generates the maximum reward. The \emph{Attack mitigating Controller Agent ($\Lambda^{c}$) } 
is activated when an attack is detected in the considered closed loop system (refer Algo.~\ref{algMultiAgentFramework}).

\begin{figure*}[!htb]
	\centering
	\begin{subfigure}[b]{0.67\columnwidth}	\includegraphics[width=\linewidth]{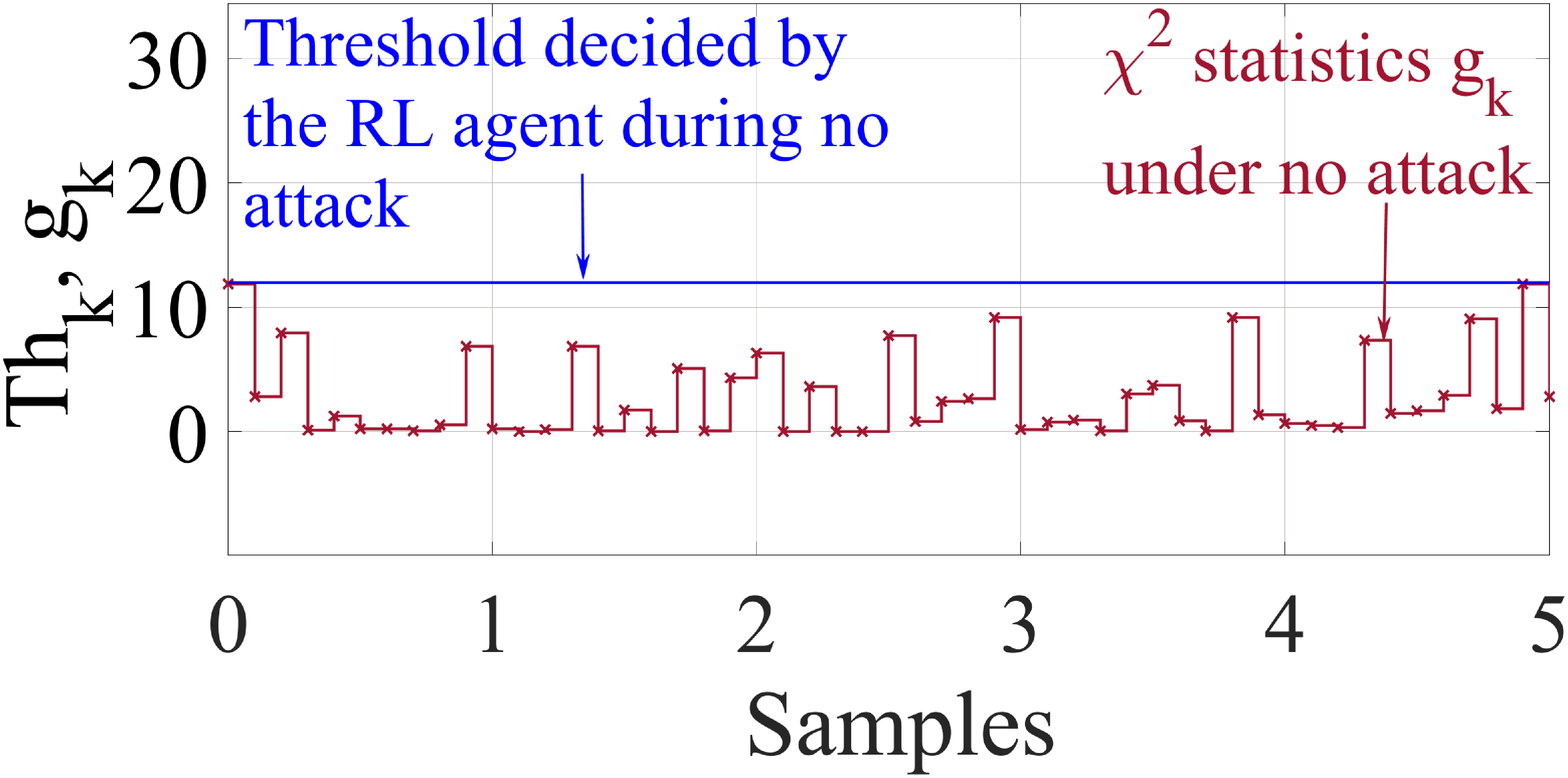}
		\caption{Threshold under no attack}
		\label{figAdaptiveNoAtk}
	\end{subfigure}
	\begin{subfigure}[b]{0.67\columnwidth}	\includegraphics[width=\linewidth]{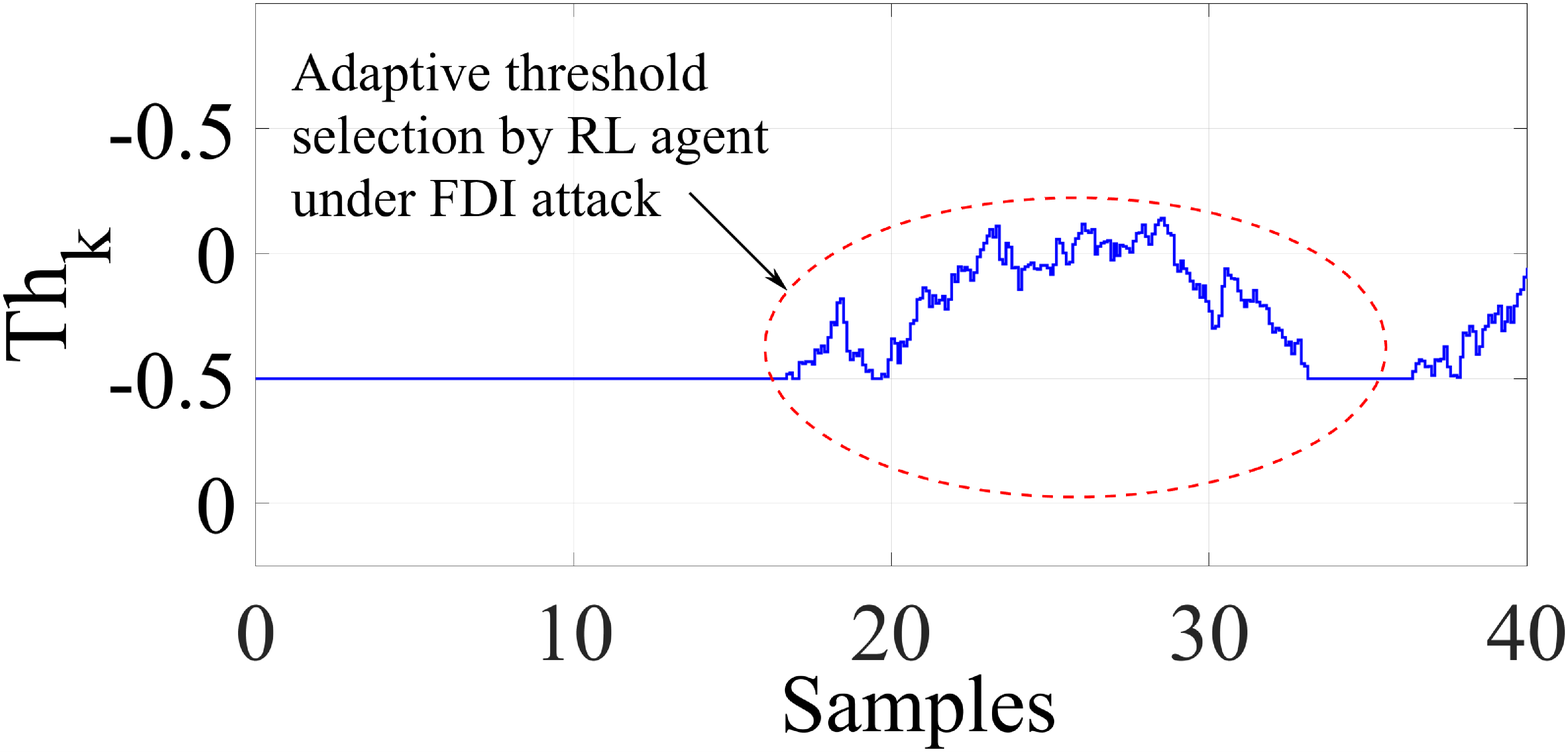}
		\caption{Variable threshold synthesis under FDI attack}
		\label{figAdaptiveAtk}
	\end{subfigure}
	\begin{subfigure}[b]{0.67\columnwidth}	\includegraphics[width=\linewidth]{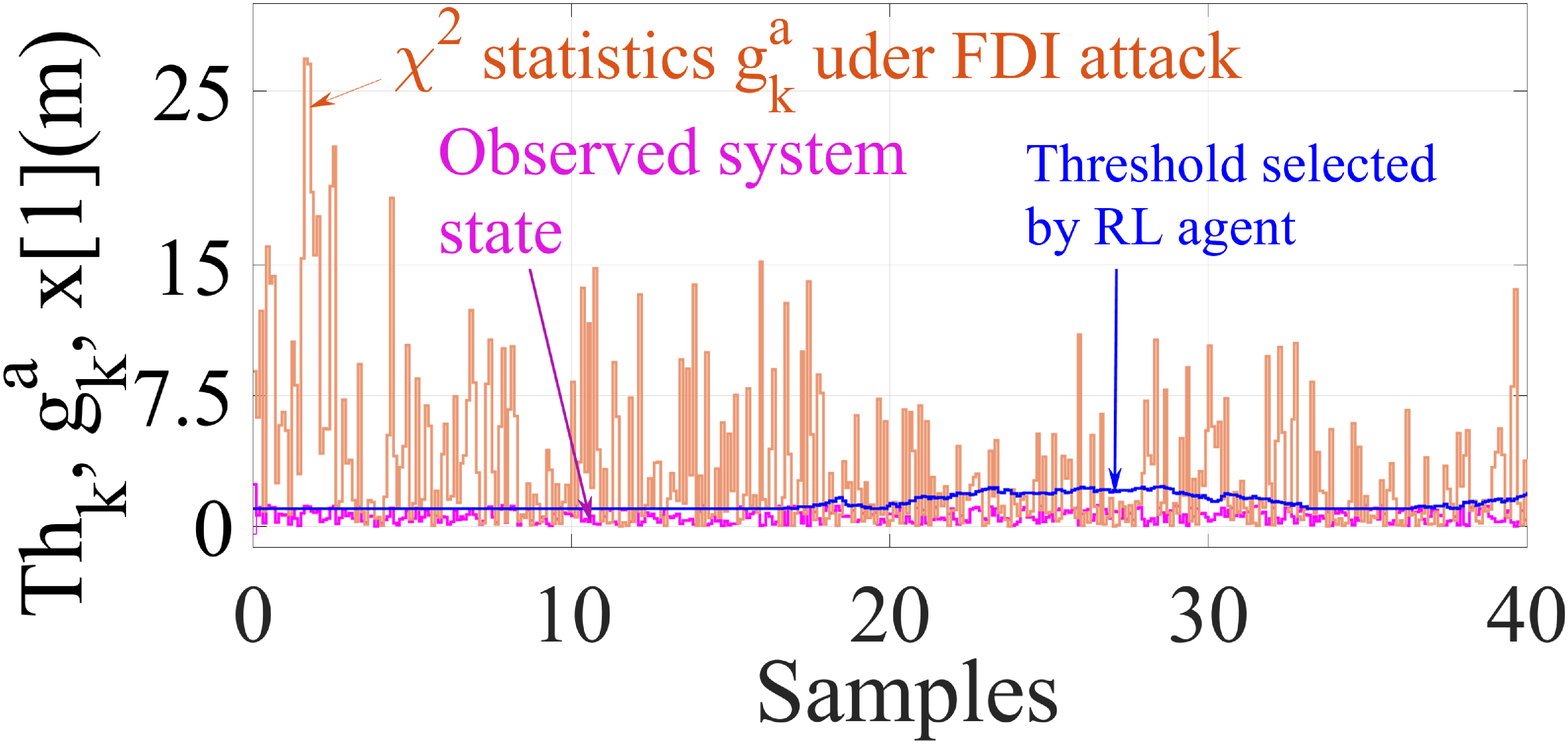}
		\caption{Attack detection}
		\label{figAdaptiveAtkAll}
	\end{subfigure}	
	\caption{Performance of adaptive threshold based detector}
	%		\vspace{-5mm}
	\label{figAdaptiveDetector}
\end{figure*}

Given a secure CPS model, we train these model-free DDPG agents so that they act according to their designated roles in the environment, which is a simulated secure CPS model.
Using standard DDPG training algorithm, how these agents ($\Lambda^a,\Lambda^d\ and\ \Lambda^c$) learn to interact with our multi-agent RL environment collaboratively and competitively, is depicted in Fig.~\ref{figRL}. For example, from the reward and objective functions we can see that the detector agent $\Lambda^d$ always tries to win against the attacker agent $\Lambda^a$ by competitively choosing an action($\chi^2$-detection threshold and observation window) to detect its ($\Lambda_a$'s) actions (FDIs). Once the attack is detected, the controller agent $\Lambda^c$ then tries to nullify the effects of these actions (attacks/FDIs output by the attacker $\Lambda^a$)  by choosing an optimal action (attack mitigating control input) from the training experiences. The attacker agent $\Lambda^a$ also tries to contest these two agents by optimally launching a \emph{hard-to-detect yet successful} FDI in every iteration. In a way, $\Lambda^a$ is collaborating with $\Lambda^d$ and $\Lambda^c$ to help them learn the system characteristics under FDI attack. Algo.~\ref{algMultiAgentFramework} represents the overall methodology.

\section{Results}
\label{secResults}

\textbf{Systems and Framework Specifications:} Automotive systems have heterogeneous communication protocols for internal communications between the Electronic Control Units (ECUs) that execute real time control tasks. 
Vulnerability in any of those protocols (eg. Controller Area Network) can grant an easy access to the attacker to manipulate majority of the system communications. This motivates us to apply our RL based monitoring framework to one such automotive CPS, namely 
Trajectory Tracking Controller (TTC). TTC regulates the deviation of a vehicle from a given trajectory ($D$) and a reference velocity ($V$) by applying proper acceleration~\cite{lesi2020integrating}. Our RL based framework is built on MATLAB Reinforcement Learning Toolbox. As mentioned earlier we employ DDPG agents having a policy gradient based actor network coupled with a DQN based critic network. Both the actor and critic networks have 3 hidden layers with rectified liner activation units (ReLU) for better  training considering the complexity of CPSs. 
We train this RL based monitoring framework with TTC for 3000 episodes with  three RL agents. The system matrices ($A,B,C$), sampling period ($h$), controller and estimator gains ($K,L$) along with corresponding preferable operating region ($X_R$) and  safety region ($X_S$) of the systems are  given in Tab.~\ref{tabSysSpecs}.
\begin{table}[!h]
\centering
%\small
\caption{System Specifications}
\label{tabSysSpecs}
\scalebox{0.8}{
\begin{tabular}{|l|l|l|l|l|}
\hline
\rowcolor[HTML]{C0C0C0} 
\multicolumn{1}{|c|}{\cellcolor[HTML]{C0C0C0}Sys.} & \multicolumn{1}{c|}{\cellcolor[HTML]{C0C0C0}Specifications} & \multicolumn{1}{c|}{\cellcolor[HTML]{C0C0C0}$X_S$} & \multicolumn{1}{c|}{\cellcolor[HTML]{C0C0C0}$X_R$} 
% &\multicolumn{1}{c|}{\cellcolor[HTML]{C0C0C0}$Th$}
\\
\hline
\begin{tabular}
[c]{@{}l@{}}TTC\end{tabular} &
\begin{tabular}
[c]{@{}l@{}}A = {[}1.0000, 0.1000;0, 1.0000{]};\\ 
B = {[}0.0050;0.1000{]}; C = {[}1 0{]};\\ 
h = 0.1sec; K = {[}16.0302,    5.6622{]};\\
L = {[}1.8721;9.6532{]}\end{tabular} & \begin{tabular}[c]{@{}l@{}}$D \in$ {[}-25, 25{]}\\
$V \in$ {[}-30, 30{]}\end{tabular} &
\begin{tabular}
[c]{@{}l@{}}$D \in$ {[}-7.5, 7.5{]}\\
$V \in$ {[}-9, 9{]}\end{tabular} 
% &2
\\
\hline
\end{tabular}
}
\end{table}

\textbf{Experimental Results: } As per our system specifications the TTC is equipped with a $\chi^2$ detector. To promise optimal resilience, while training, we assume the attacker is aware of  the adaptive threshold based detector specifications (the currently chosen $Th_k$ and $l$). We have trained the detector agent in the presence and absence of the attacker agent to reinforce the learning that differentiates an attacked and un-attacked situation. 
Given the safety specifications of the system we first derive the preferable operating region of TTC using Algo~.\ref{algPrefOptReg}. Intializing the system states from this region, we train the RL agents for 3000 episodes, each with 100 simulation instances in order to train them. 
As we can see in Fig.~\ref{figAdaptiveAtk} , the detector agent explores and exploits different threshold values for different detection windows. This is an FDI attack scenario where the attacker agent injects optimal false data to make the system unsafe. For this, the $\chi^2$-test value on the system residue changes as we see in Fig.~\ref{figAdaptiveAtkAll}. The detector agent starts by selecting $Th=1$ for $l=4$ and changes the threshold to successfully detect most of the attack efforts. If we consider the maximum non-centrality induced by the optimal FDI attacker agent, TPR achieved by the adaptive detector agent is $0.91$. Note that, as shown in Fig.~\ref{figAdaptiveAtkAll} under the considered optimal attack scenario, the attacker is detected even before it is able to send the observed system state outside the preferable operating region $X_R$ (=7.5 for the output state, refer Tab.~\ref{tabSysSpecs}). Hence the robust controller does not kick off and our adaptive detection system promises a cost effective control by early detection of attack.

In Fig.~\ref{figAdaptiveNoAtk} as can be seen, our variable threshold based detector selects $Th=12$ when there is no FDI attack in place. The average FAR achieved by the designed adaptive detection system during no attack situation is $0.04$. As we can observe in Fig.~\ref{figAdaptiveNoAtk}, it manages to place the threshold above the $\chi^2$-test values of system residues due to noises (under no FDI attack). Targeting the minimum FAR, consider a constant threshold based $\chi^2$ detector with $Th=12$ is placed to detect FDI attacks. Then, it is clear from the $\chi^2$ statistics of residues under attack in Fig.~\ref{figAdaptiveAtkAll} that many of the attack attempts would have remained undetected.

	\section{Conclusion} The present work  proposed a  RL based secure CPS model and studied its  usefulness through simulation using an automotive CPS benchmark. In future, we plan to 
	create an automotive hardware-in-loop simulation infrastructure which will help us simulate automotive control loops and vehicle dynamics in a real time platform in order to  
	check the timing performance of the proposed scheme in a more realistic setting.

	\bibliographystyle{IEEEtran}
	\bibliography{reference.bib}

\end{document}